\documentclass[aps,prl,twocolumn,superscriptaddress,longbibliography]{revtex4-2}
\usepackage{graphicx}
\usepackage{amssymb,amsmath}
\usepackage{bm}
\usepackage{dcolumn}
\usepackage{float}
\usepackage[OT1]{fontenc} 
\usepackage{url}
\usepackage{mathrsfs}
\usepackage{slashed,comment}
\usepackage{color}
\usepackage{verbatim}
\usepackage{enumitem}
\usepackage{soul,physics}
\usepackage[driverfallback=dvipdfm]{hyperref}
\hypersetup{pdfpagemode=FullScreen,colorlinks=true,breaklinks,urlcolor=blue,linkcolor=blue,citecolor=blue}

\usepackage{subfigure}
\usepackage{amssymb}
\usepackage{bm}
\usepackage{graphicx}
\usepackage{amsmath,amssymb}

\usepackage{pdfpages} 
\usepackage{pgffor} 

\makeatletter
\AtBeginDocument{\let\LS@rot\@undefined}
\makeatother

\def\supplementfilename{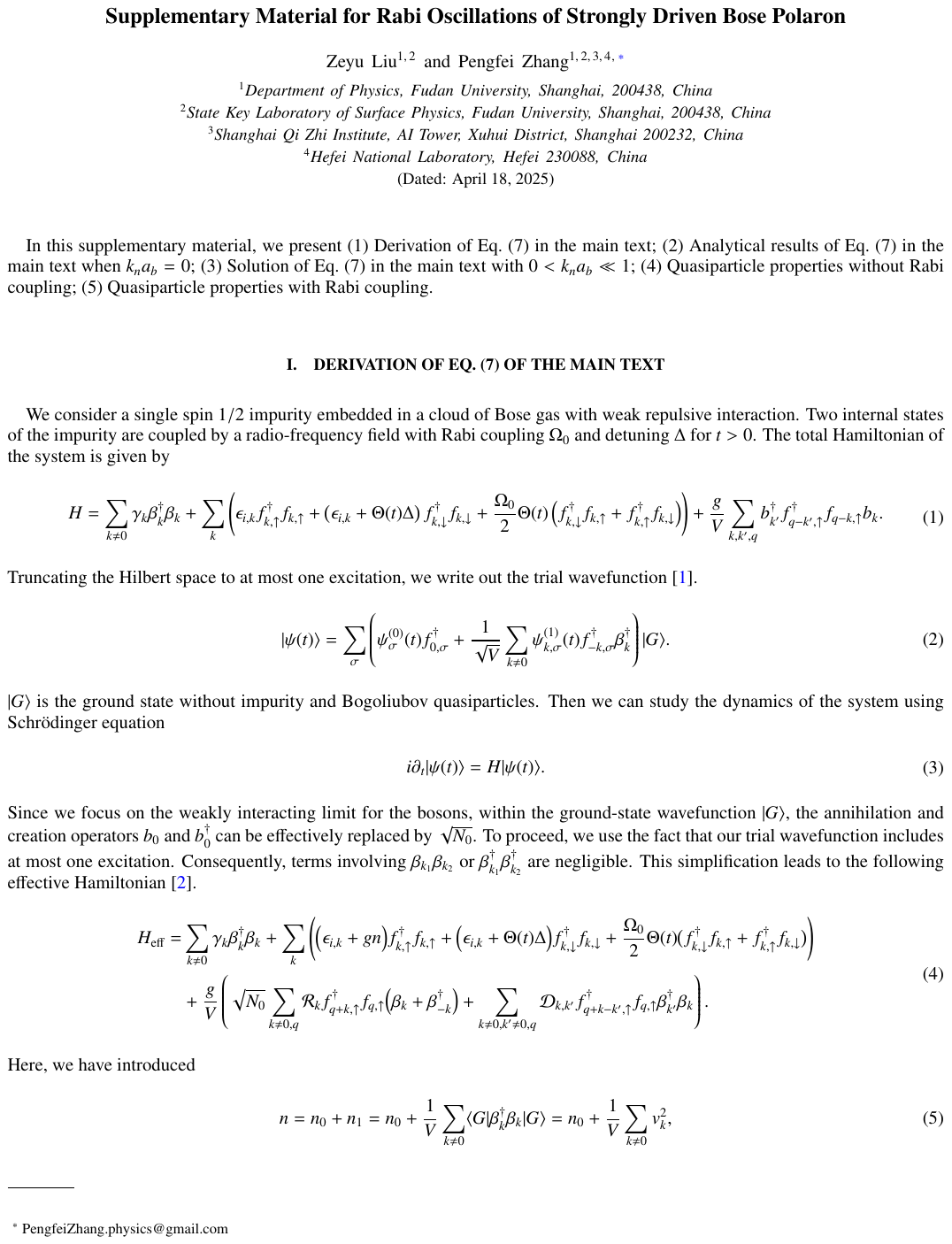}

\pdfximage{\supplementfilename}
\def\numbersupplementpages{\the\pdflastximagepages}

\newif\ifarXiv
\arXivtrue 

\begin{document}
 
  \title{Rabi Oscillations of Strongly Driven Bose Polarons }

  \author{Zeyu Liu}
  \affiliation{Department of Physics, Fudan University, Shanghai, 200438, China}
  \affiliation{State Key Laboratory of Surface Physics, Fudan University, Shanghai, 200438, China}

  \author{Pengfei Zhang}
  \thanks{PengfeiZhang.physics@gmail.com}
  \affiliation{Department of Physics, Fudan University, Shanghai, 200438, China}
  \affiliation{State Key Laboratory of Surface Physics, Fudan University, Shanghai, 200438, China}
  \affiliation{Shanghai Qi Zhi Institute, AI Tower, Xuhui District, Shanghai 200232, China}
  \affiliation{Hefei National Laboratory, Hefei 230088, China}

  \date{\today}

  \begin{abstract}
  Understanding the dynamical behavior of quasiparticles is essential for uncovering novel quantum many-body phenomena. Among these phenomena, the polaron in ultracold atomic gases has attracted considerable interest due to its precise controllability. By engineering the underlying Hamiltonian, polarons serve as a versatile platform for studying both equilibrium properties and non-equilibrium dynamics. In this work, we investigate the quantum dynamics of strongly driven Bose polarons, where minority atoms are described as mobile impurities with spin-1/2, and majority atoms are bosons. The spin-$\uparrow$ impurity interacts with majority atoms through a tunable scattering length $a$, while the spin-$\downarrow$ impurity remains non-interacting. After turning on the Rabi coupling, we calculate the evolution of the total magnetization using a trial wavefunction. We identify the exhibition of anomalous Rabi oscillation and steady-state magnetization for $a>0$, due to the interplay between attractive and repulsive polarons. Our results provide a concrete example that illustrates how Rabi oscillations are dressed by system-environment coupling.
  \end{abstract}
    
  \maketitle

  \emph{ \color{blue}Introduction.--} 
  The emergence of quasiparticles is a cornerstone of many-body physics, providing an effective description of complex interactions through renormalized excitations. From Landau’s Fermi liquid theory to fractionalized excitations in topological phases, quasiparticles serve as a powerful conceptual framework for describing collective behavior. Among these, the polaron, an impurity dressed by excitations of a surrounding medium, plays an important role in understanding the properties of various physical systems, ranging from solid-state materials to ultracold atomic gases \cite{Landau:1933asb,Landau:1948ijj,Landau:1956yop}. In ultracold atomic gases, the setup consists of minority atoms immersed in a cloud of majority atoms at low temperatures. The interaction between the minority atom and the majority atoms can be tuned using the Feshbach resonance \cite{Chin:2010crf}. Depending on the statistics of the majority atoms, the polaron is classified as either a Bose polaron \cite{astrakharchik2004motion,Sacha:2006vfafa,Cucchietti:2006bas,Bruderer:2008,Tempere:2009bva,Huang:2009,Catani:2012vwa,Spethmann:2012bss,Scelle:2013vrds,balewski:2013crx,Blinova:2013new,Rath:2013brtd,Li:2014var,Marti:2014ves,Shashi:2014vsr,ardila:2015im,Christensen:2015vas,Grusdt:2015uena,Hu:2016vaa,Jorgensen:2016brw,Shchadilova:2016vasryt,Levinsen:2017hvae,Yoshida:2018bva,Yoshida:2018bvs,Grusdt:2017yrna,Guenther:2018ure,Yan:2020gag,Field:2020usaj,Drescher:2019eif,Drescher:2020edca,Skou:2021bose,Levinsen:2021bbr,Massignan:2021nde,Skou:2022bose,Cayla:2023unia,schmidt:2022vesad,Mostaan:2023brads,Vivanco:2025fermi,Morgen:2025trfs,Etrych:2025inu,Levinsen:2015baef,Levinsen:2024brdf,Nakano:2024vea,Christianen:2024baz} or a Fermi polaron \cite{Chevy:2006,Lobo:2006brr,Combescot:2007vae,prokofev:2008buw,Combescot:2008rsc,Pilati:2008bnt,nascimbene:2009brs,Schirotzek:2009vsrt,cui:2020aev,Kohstall:2012vbae,Koschorreck:2012aje,Zhang:2012vsr,Wenz:2013ayr,Cetina:2016vaa,Li:2017vrs,Scazza:2017vrs,Liu:2019bes,Yan:2019fva,Darkwah:2019vge,Ness:2020faw,cui:2010aev,Peng:2021fbt,mulkerin:2024fsad,baroni:2024vwa}. A Bose polaron forms when minority atoms are dressed by phonon excitations, whereas a Fermi polaron arises from the excitation of particle-hole pairs. Studies also highlight the signature of Efimov physics in the Bose polaron with large mass imbalance \cite{Efimov:1970zz,Kraemer:2006crx,Pires:2014btd,Huang:2014yan,Tung:2014vstr,Levinsen:2015baef,Shi:2015vas,sun:2017bss,Zulli:2025}.

  \begin{figure}[t]
    \centering
    \includegraphics[width=0.75\linewidth]{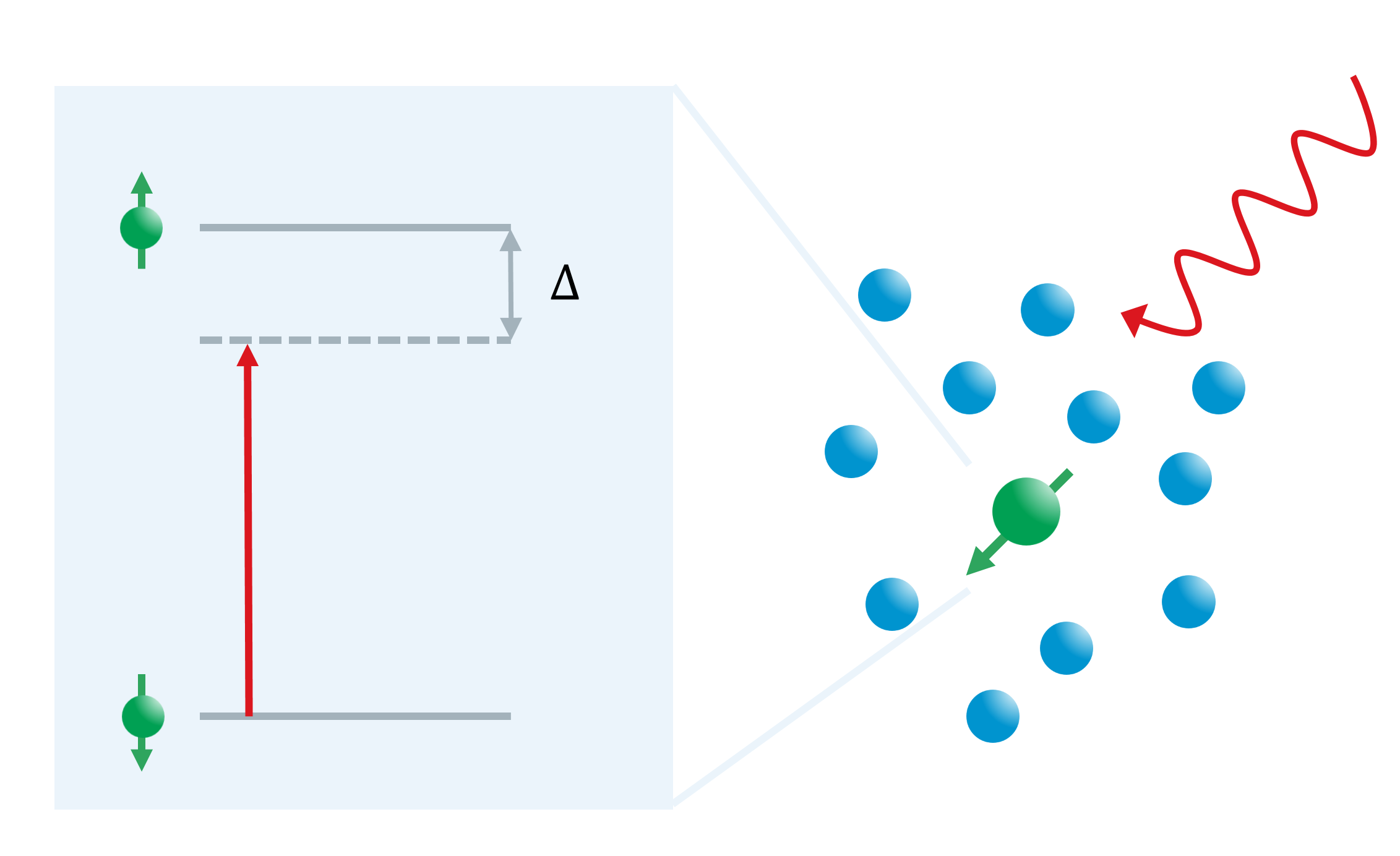}
    \caption{We present a schematic of the setup of our system. We consider an impurity atom (green) immersed in a cloud of bosonic atoms (blue). The impurity has two internal states, labeled by $\uparrow$ and $\downarrow$. Initially, the impurity is prepared in the state $\ket{\downarrow}$, while the interaction between the impurity and the environment is only present for $\ket{\uparrow}$. We study the dressed Rabi oscillation of the system after a strong radio-frequency field is turned on, using the trial wavefunction approach.  }
    \label{fig:schemticas}
  \end{figure}

  Most studies on Bose/Fermi polarons focus on spectral functions near thermal equilibrium, typically measured using ejection and injection spectroscopy. Experimentally, this requires weak coupling of the internal state of the minority atoms to an auxiliary state, where the interaction between the minority and majority atoms is negligible. After a short time, the population of the minority atoms is measured, following Fermi's Golden Rule \cite{Schirotzek:2009vsrt,Kohstall:2012vbae,Koschorreck:2012aje,Hu:2016vaa,Jorgensen:2016brw,zhai2021ultracold}. Recently, there has been growing interest in exploring the opposite regime, where a strong Rabi coupling is applied for a sufficiently long time, allowing the system to reach a steady state \cite{Vivanco:2025fermi}. On the one hand, this regime investigates the dressing of Rabi oscillations in the presence of system-environment coupling, with a close relation to quantum thermalization. On the other hand, it provides a concrete example of the fate of quasiparticles under a strong driving field. For Fermi polarons, theoretical calculations near unitarity using the Green's function approach have shown good agreement with experimental results \cite{mulkerin:2024fsad}. However, an analysis of strongly driven Bose polarons is still lacking.
  
  In this work, we take the first step toward understanding the Rabi oscillations of strongly driven Bose polarons by formulating a trial wavefunction approach \cite{Chevy:2006} to investigate their non-equilibrium dynamics. When the majority atoms are non-interacting, our approach allows for an analytical treatment, providing an accurate description of the dynamical evolution for arbitrarily long times. The results reveal the emergence of anomalous Rabi oscillations and steady-state magnetization on the BEC side, where the attractive and repulsive branches coexist. We explain these phenomena by relating the Fourier transform of the impurity polarization density over time to the steady-state spectral functions, which can be approximated using perturbation theory. Furthermore, we provide numerical evidence that introducing a weak interaction between majority atoms does not lead to a qualitative change in our conclusions. We expect that our theoretical prediction can be readily tested on state-of-the-art experimental platforms.

  \emph{ \color{blue}Model.--} 
  Our model consists of a single spin-1/2 impurity immersed in a cloud of bosonic atoms, as illustrated in Fig. \ref{fig:schemticas}. The bosons are either non-interacting or have weak repulsive interactions. Both cases can be captured by the Bogoliubov Hamiltonian
  \begin{equation}
  \begin{aligned}
  H_{b}=\sum_{k\neq 0}\gamma_k\beta_{k}^{\dagger}\beta_{k}. 
  \end{aligned}
  \end{equation}
  Here, $\gamma_k=\sqrt{\epsilon_{b,k}\Big( \epsilon_{b,k}+2g_{b}n_{0} \Big)}$ is the dispersion of Bogoliubov quasiparticles. We define $\epsilon_{b,k}={k^{2}}/{2m_{b}}$ as the dispersion for non-interacting bosons, and $g_b$ is related to the scattering length $a_b$ between bosons by the relation $g_b={4\pi a_{b}}/{m_{b}}$. The annihilation operator of the Bogoliubov quasiparticle $\beta_{k}$ is related to the annihilation and creation operators of atoms $b_{k}/b_{-k}^\dagger$ via the transformation $\beta_{k}=u_{k}b_{k}+v_{k}b_{-k}^{\dagger}$, where $u_{k}^{2}=\frac{1}{2}\Big( \frac{\epsilon_{b,k}+g_{b}n_{0}}{\gamma_k}+1 \Big)$ and $u_k^2-v_k^2=1$. The impurity has two internal states, $\ket{\uparrow}$ and $\ket{\downarrow}$, which are coupled through a radio-frequency field with Rabi coupling $\Omega_0$ and detuning $\Delta$ for $t>0$. The Hamiltonian of the impurity reads
  \begin{equation}
  \begin{aligned}
  H_i(t)=\sum_{k}\Big( &\epsilon_{i,k}f_{k,\uparrow}^{\dagger}f_{k,\uparrow}+\big(  \epsilon_{i,k}+{\Theta(t)}\Delta \big)f_{k,\downarrow}^{\dagger}f_{k,\downarrow}\\&+\frac{\Omega_{0}}{2}\Theta(t)\big(f_{k,\downarrow}^{\dagger}f_{k,\uparrow}+f_{k,\uparrow}^{\dagger}f_{k,\downarrow}\big) \Big).
  \end{aligned}
  \end{equation} 
  Here, $f_{k,\sigma}$ annihilates an impurity atom with spin $\sigma\in \{\uparrow,\downarrow\}$ and $\epsilon_{i,k}={k^{2}}/{2m_{i}}$. $\Theta(t)$ is the Heaviside step function. Since we focus on the single-impurity sector, the statistics of the impurity are irrelevant to our analysis. Nevertheless, we assume the impurity is fermionic {for} concreteness. Finally, the impurity interacts with the bosons only if it is in the state $\ket{\uparrow}$. This leads to
  \begin{equation}
  \begin{aligned}
  H_{bi}=\frac{g}{V}\sum_{k,k^{\prime},q}b_{k^{\prime}}^{\dagger}f_{q-k^{\prime},\uparrow}^{\dagger}f_{q-k,\uparrow}b_{k}.
  \end{aligned}
  \end{equation}
  The interaction term has an ultraviolet divergence, which is regularized by the renormalization relation $\frac{1}{g}=\frac{m}{2\pi a}-\frac{1}{V}\sum_{k}\frac{2m}{k^{2}}$, where $m^{-1}=m_{b}^{-1}+m_{i}^{-1}$ is the inverse reduced mass. The Hamiltonian of the total system is therefore given by $H(t)=H_b+H_i(t)+H_{bi}$.

 \begin{figure}[t]
    \centering
    \includegraphics[width=0.99\linewidth]{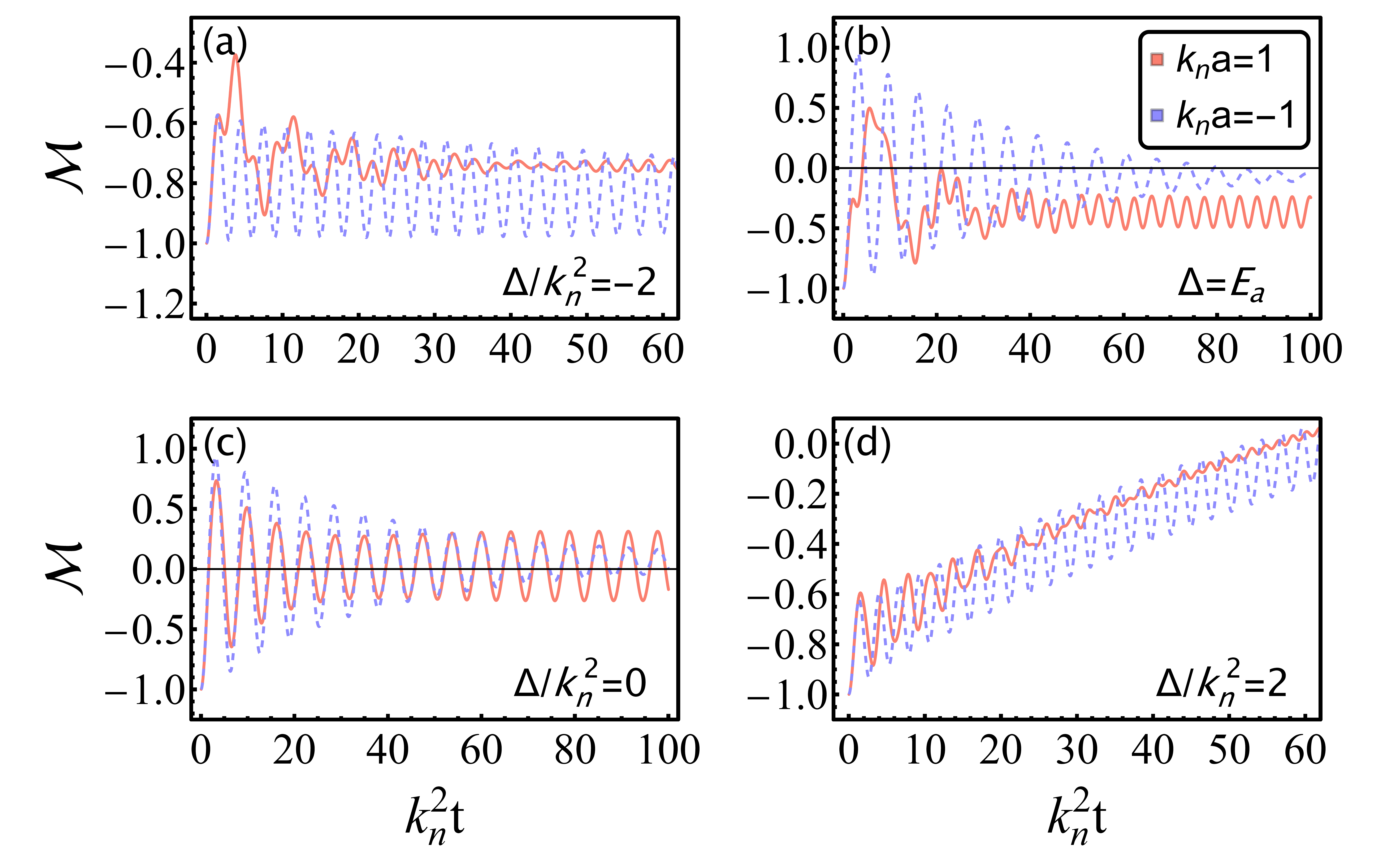}
    \caption{We present numerical results for the magnetization dynamics during Rabi oscillations with $k_{n}a_{b}=0$ and $\Omega_{0}/k_{n}^{2}=1$ for different impurity-boson interactions with $k_na =\pm 1$. The analysis includes four different detunings: $\Delta=-2k_n^2,E_a, 0, 2k_n^2$, where $E_a$ represents the energy of the attractive polaron (without the Rabi coupling), which satisfies $E_{a}=\Sigma(E_{a})$. For $k_n a=1$, this leads to $\Delta\approx-1.34k_n^2$.  }
    \label{fig2}
  \end{figure}

  \emph{ \color{blue}Trial wavefunction.--} 
  For $t<0$, the impurity is prepared in the spin-$\downarrow$ state with zero momentum, and the majority atoms are in the ground state with no quasiparticle excitation. This corresponds to a product state 
  \begin{equation}
  |\psi(0)\rangle=f^\dagger_{0,\downarrow}|0\rangle_i|\text{BEC}\rangle_b\equiv f^\dagger_{0,\downarrow}|G\rangle,
  \end{equation}
  where $\beta_k|\text{BEC}\rangle_b=0$. At $t=0$, the Rabi coupling is turned on and the system begins to evolve. In particular, after the transition into the spin-$\uparrow$ state, the impurity-boson interaction can excite Bogoliubov quasiparticles. Truncating the Hilbert space to at most one excitation, we write out the trial wavefunction.
  \begin{equation}
  \begin{aligned}
  |\psi(t)\rangle=\sum_{\sigma}\Big( \psi_{\sigma}^{(0)}(t)f_{0,\sigma}^{\dagger}+\frac{1}{\sqrt{V}}\sum_{k}\psi_{k,\sigma}^{(1)}(t)f_{-k,\sigma}^{\dagger}\beta_{k}^{\dagger} \Big)|G\rangle.
  \end{aligned}
  \end{equation}
  Here, $V$ is the size of the total system. The summation over $k$ implicitly excludes $k=0$. The factor of ${1}/{\sqrt{V}}$ is introduced such that wavefunctions satisfy the normalization condition 
  \begin{equation}
  \sum_\sigma \Bigg[|\psi_{\sigma}^{(0)}(t)|^2+\int \frac{d\mathbf{k}}{(2\pi)^3}|\psi_{k,\sigma}^{(1)}(t)|^2\Bigg]=1.
  \end{equation}
  Similar trial wavefunctions have been employed to study both the equilibrium properties and quench dynamics of Bose polarons in the absence of an external drive \cite{Li:2014var,Yoshida:2018bvs,Yoshida:2018bva,Levinsen:2021bbr}. It serves as a good approximation across the Feshbach resonance, capturing the crossover between the polaron state and the molecule state for the attractive branch, as well as the presence of a repulsive branch for $a>0$. It is also straightforward to include contributions from two-excitation sectors, which contain three-body correlations. This is particularly important for Bose polarons with a high mass imbalance. In this work, we instead focus on the simplest scenario with $m_b=m_i$, which is realized by using different hyperfine states of the same atomic species for both the impurity and the bosonic environment. 

  Leaving the details to the supplementary material \cite{SM}, for a generic scattering length $a_b$, the evolution of the wavefunction reads
  \begin{equation}\label{eqn:evolution_eqn}
  \begin{aligned}
  i\partial_{t}\psi_{\uparrow}^{(0)}=&gn_0\psi_{\uparrow}^{(0)}+\frac{\Omega_{0}}{2}\psi_{\downarrow}^{(0)}+\frac{g}{V}\sqrt{n_{0}}\sum_{k}\mathcal{R}_k\psi_{k,\uparrow}^{(1)},\\
  i\partial_{t}\psi_{k,\uparrow}^{(1)}=&\Big( \epsilon_k+gn_0 \Big)\psi_{k,\uparrow}^{(1)}+\frac{\Omega_{0}}{2}\psi_{k,\downarrow}^{(1)}\\&+g\sqrt{n_{0}}\mathcal{R}_k\psi_{\uparrow}^{(0)}+\frac{g}{V}\sum_{k^{\prime}}\mathcal{D}_{k,k^{\prime}}\psi_{k^{\prime},\uparrow}^{(1)},\\
  i\partial_{t}\psi_{\downarrow}^{(0)}=&\Delta\psi_{\downarrow}^{(0)}+\frac{\Omega_{0}}{2}\psi_{\uparrow}^{(0)},\\
  i\partial_{t}\psi_{k,\downarrow}^{(1)}=&\Big( \epsilon_k+\Delta \Big)\psi_{k,\downarrow}^{(1)}+\frac{\Omega_{0}}{2}\psi_{k,\uparrow}^{(1)}.
  \end{aligned}
  \end{equation}
  Here, we have introduced $\epsilon_{k}=\epsilon_{i,k}+\gamma_{k}$, $\mathcal{R}_k=u_k-v_k$, and $\mathcal{D}_{k,k^{\prime}}=u_ku_{k'}+v_kv_{k'}$ for conciseness. $n_0$ is the density of the condensate. Since we focus on a weakly interacting or non-interacting Bose gas, we do not distinguish $n_0$ from the total density of the majority atoms. The initial condition is $\psi^{(0)}_{\downarrow}=1$, while all other components of the wavefunction vanish. In the following sections, we analyze this set of equations to determine the dynamics of strongly driven Bose polarons. In particular, we focus on the magnetization of the impurity, defined as
  \begin{equation}\label{eqn:mag}
  \begin{aligned}
  \mathcal{M}(t)=&|\psi_{\uparrow}^{(0)}(t)|^2-|\psi_{\downarrow}^{(0)}(t)|^2\\&+\int \frac{d\mathbf{k}}{(2\pi)^3}(|\psi_{k,\uparrow}^{(1)}(t)|^2-|\psi_{k,\downarrow}^{(1)}(t)|^2).
  \end{aligned}
  \end{equation}

    \begin{figure}[t]
    \centering
    \includegraphics[width=0.99\linewidth]{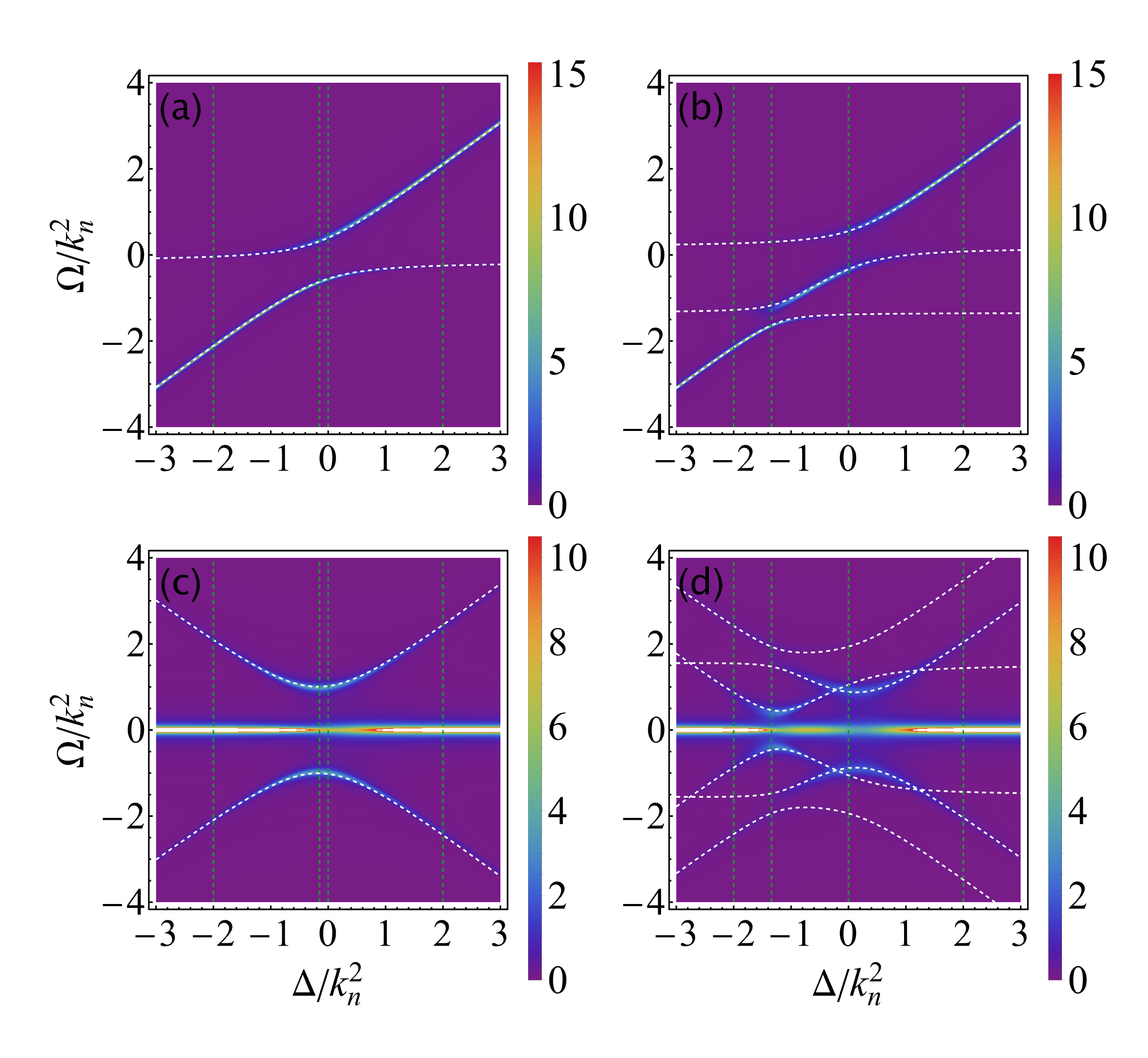}
    \caption{We present the density plot for the impurity spectral function $A_{\downarrow}(\Omega)$ (a-b) and the Rabi spectrum (c-d). Here, we choose $k_{n}a = -1$ for panels (a) and (c), and $k_{n}a = 1$ for panels (b) and (d). In all plots, we set $k_{n}a_{b}=0$ and $\Omega_{0}/k_{n}^{2}=1$. Vertical dashed lines indicate the choices of $\Delta$ for FIG. \ref{fig2}. White dashed lines show the theoretical predictions from the perturbative calculation, which match the numerical results with good accuracy.  }
    \label{fig3}
  \end{figure}

  \emph{ \color{blue}Results.--} We first focus on the limit of $a_b=0$, where bosons are non-interacting. The Eq. \eqref{eqn:evolution_eqn} is greatly simplified since $u_k=1$ and $v_k=0$. After performing the Laplace transform of the wavefunction $f(s)\equiv \int_{0}^{\infty} e^{-st}f(t)~dt$, we can obtain analytical expressions for all components of the wavefunction. The central observation is that we have $\frac{g}{V}\sum_k \mathcal{R}_k\psi_{k,\uparrow}^{(1)}=\frac{g}{V}\sum_k \mathcal{D}_{k',k}\psi_{k,\uparrow}^{(1)}=\frac{g}{V}\sum_k \psi_{k,\uparrow}^{(1)}$. Therefore, Eq. \eqref{eqn:evolution_eqn} allows all components of the wavefunction in terms of $\frac{g}{V}\sum_k \psi_{k,\uparrow}^{(1)}$. For example, this leads to the relation
   \begin{equation}\label{eqn:e2}
  \begin{aligned}
  \psi_{\uparrow}^{(0)}(s)&=\frac{\frac{i\Omega_{0}}{2}+{(is-\Delta)\sqrt{n_0}\chi(s)}}{D_0},\\
  \psi_{k,\uparrow}^{(1)}(s)&=\frac{(is-\epsilon_k-\Delta)\chi(s)}{D_k},
  \end{aligned}
  \end{equation}
  where, for conciseness, we define $D_k=(is-\epsilon_k)(is-\epsilon_k-\Delta)-\left( \frac{\Omega_{0}}{2} \right)^{2}$. We also introduce $\chi(s)=g\sqrt{n_0}\psi_{\uparrow}^{(0)}(s)+\frac{g}{V}\sum_k \psi_{k,\uparrow}^{(1)}(s)$. Combining this definition with Eq. $\eqref{eqn:e2}$, we derive the result for $\chi(s)$ as
  \begin{equation}
  \begin{aligned}
  &\left(\Pi(is)-\frac{(is-\Delta)n}{D_0}\right)\chi(s)=\frac{i\Omega_{0}\sqrt{n}}{2D_0},\\&\Pi(is)=\frac{m}{2\pi a}-\int \frac{d\mathbf{k}}{(2\pi)^3}\left[\frac{is-\epsilon(k)-\Delta}{D_{k}}+\frac{2m}{k^{2}}\right],
  \end{aligned}
  \end{equation}
  where we have used the renormalization relation to absorb the divergence in $\Pi({i}s)$. In the limit of $\Omega_0=0$, $\Pi(E)|_{\Omega_{0}=0}\equiv \Sigma(E)$ matches (the inverse of) the two-body $T$-matrix. The integration over $\mathbf{k}$ can be performed analytically for arbitrary $\Omega_0$, and its explicit form is provided in the supplementary material \cite{SM}. This leads to a closed-form expression for both $\psi_{\sigma}^{(0)}(s)$ and $\psi_{k,\sigma}^{(1)}(s)$. However, we are not able to perform the inverse Laplace transform analytically. Therefore, we numerically determine $\psi_{\sigma}^{(0)}(t)$ and $\psi_{k,\sigma}^{(1)}(t)$ (for a discretized set of $k$ values) and then compute the magnetization using Eq. \eqref{eqn:mag}. 

   \begin{figure}[t]
    \centering
    \includegraphics[width=0.75\linewidth]{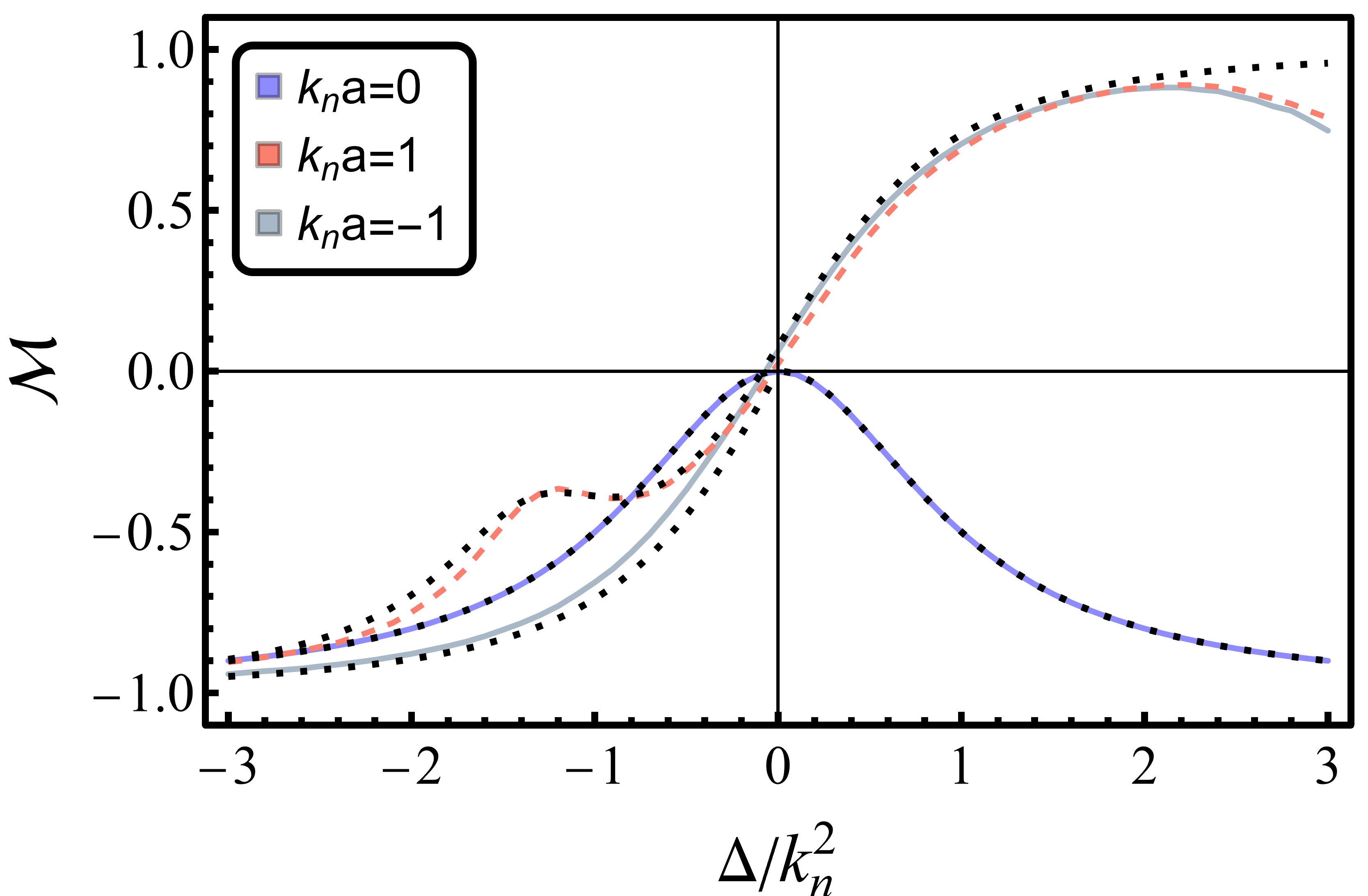}
    \caption{We present the steady-state magnetization as a function of detuning $\Delta$ for $k_na=0,1,-1$, with fixed parameters $k_{n}a_{b}=0$ and $\Omega_{0}/k_{n}^{2}=1$. A prominent peak near $\Delta\approx -1.34 k_n^2\approx E_a$ serves as a signature of the attractive polaron (see discussions in the main text). The decrease for $k_n a=\pm 1$ at large $\Delta$ arises from the insufficient evolution time to fully reach the steady state. }
    \label{fig4}
  \end{figure}

  \textbf{ Short-time oscillation.--}Numerical results in both the BEC ($k_n a=1$) and BCS ($k_n a=-1$) regimes, for different values of detuning $\Delta$, are presented in FIG. \ref{fig2}, where the density is parametrized as $n_{0}={k_{n}^{3}}/{6\pi^{2}}$. We summarize several interesting features of the magnetization dynamics. At short times, we observe many-body Rabi oscillations with a damped amplitude due to coupling with the environment. For either detuning $\Delta \gtrsim 0$ or $k_n a =-1$, the oscillation is regular and approximately single-frequency. This is a close analog of Rabi oscillations of non-interacting spin-1/2 particles. In contrast, the oscillation becomes highly irregular for $k_n a = 1$ and $\Delta/k_n^2 \in \{-2, -1.34\}$, indicating a large deviation from the non-interacting case. We refer to this phenomenon as anomalous Rabi oscillation. 

  To uncover the underlying physical origin, we study the impurity spectral function in the presence of the Rabi coupling. This is done by analyzing the eigenenergy of Eq. \eqref{eqn:evolution_eqn}, replacing $i\partial_t$ with $E$. Straightforward calculations yield $E=\Delta+\frac{(\Omega_{0}/2)^{2}}{E-n\Pi^{-1}(E)}$. This equation should be interpreted as looking for the pole of the impurity Green's function 
  \begin{equation}
\begin{aligned}
  G_{\downarrow}^R(\Omega)=\frac{1}{\Omega-\Delta-\frac{(\Omega_{0}/2)^{2}}{\Omega-n\Pi^{-1}(\Omega+i0^+)}}.
\end{aligned}
\end{equation}
Here, we have introduced the retarded Green's function $G_{\downarrow}^R(t)\equiv -i\theta(t)\langle \{f_{0,\downarrow}(t),f_{0,\downarrow}^\dagger(0)\}\rangle$. Then, we can compute the spectral function using $A_{\downarrow}(\Omega)=-\frac{1}{\pi}\text{Im} \big( G_{\downarrow}^R(\Omega) \big)$. The numerical results for $A_{\downarrow}(\Omega)$ is presented in FIG. \ref{fig3} (a-b) for $k_na=-1$ and $k_na=1$, respectively. For $k_na<0$, there are two branches of excitations, formed due to the avoid-crossing between the attractive polaron formed by the spin-$\uparrow$ state and the non-interacting spin-$\downarrow$ state. This is well captured by a perturbative analysis of $\Omega_0$, which makes the single-pole approximation $\frac{1}{\Omega-n\Pi^{-1}(\Omega)}\approx \frac{Z_a}{\Omega-E_a}$. Here, the polaron energy $E_a$ and the quasi-particle residue $Z_a$ are determined by solving $E_{a}=\Sigma(E_{a})$ and $Z_{a}=\left(1-\partial_{\Omega}\Sigma(E_a)\right)^{-1}$. Under this approximation, the excitation energies $\Omega^*$ are given by:
\begin{equation}\label{eqn:twolevel}
\begin{aligned}
  \Omega^*=\frac{E_{a}+\Delta\pm\sqrt{(E_{a}-\Delta)^{2}+Z_{a}\Omega^{2}_{0}}}{2}\ \ \ \text{for}\ a<0,
\end{aligned}
\end{equation}
which are plotted in FIG. \ref{fig3}(a) using white dashed lines. A similar relation has been discovered for driven Fermi polarons \cite{mulkerin:2024fsad}. For comparison, there are three branches of excitations for $k_n a>0$ due to the coexistence of attractive and repulsive polarons \cite{cui:2010aev}. Therefore, we approximate $\frac{1}{\Omega-n\Pi^{-1}(\Omega)}\approx \frac{Z_a}{\Omega-E_a}+\frac{1-Z_a}{\Omega-E_r}$, where $E_r$ is the energy of the repulsive polaron, satisfying $E_r=\text{Re}[\Sigma(E_r)]$. This leads to the equation for excitation energies
\begin{equation}\label{eqn:threelevel}
\begin{aligned}
  (\Omega^*-\Delta)=\frac{(\Omega_{0}/2)^{2}Z_{a}}{\Omega^*-E_{a}}+\frac{(\Omega_{0}/2)^{2}(1-Z_{a})}{\Omega^*-E_{r}}\ \ \ \text{for}\ a>0.
\end{aligned}
\end{equation}
Finding explicit solutions is possible, though tedious. We plot the results in FIG. \ref{fig3}(b) using white dashed lines, which also match the numerical results to good accuracy. 

To proceed, we consider the magnetization dynamics. We perform a Fourier transform of $N_{\downarrow}^{(0)}\equiv|\psi_{\downarrow}^{(0)}(t)|^{2}$, which is referred to as the Rabi spectrum. By definition, $N_{\downarrow}^{(0)}$ is quadratic in the wavefunction. Therefore, it is natural to expect that the spectrum is determined by the energy differences between excitation energies $\Omega^*$. This prediction is demonstrated in FIG. \ref{fig3}(c-d). For $a<0$, the two excitation energies in \eqref{eqn:twolevel} lead to a single peak at $\Omega >0$. As a result, the magnetization is nearly single-frequency. For $a>0$, the coexistence of attractive and repulsive polarons gives rise to a complex Rabi spectrum. When $\Delta$ is negative but has a small magnitude, we find a finite spectral weight in all branches. Moreover, the hybridization between the attractive polaron and the spin-$\downarrow$ state becomes significant due to the renormalization of the Rabi coupling. These effects lead to the observed anomalous Rabi oscillations in the magnetization, exhibiting a sharp contrast with non-interacting spins.

\textbf{ Steady-state magnetization.--} 
We now examine the long-time behavior of many-body Rabi oscillations, where the system approaches its steady state. The steady-state magnetization is defined as $\overline{\mathcal{M}}\equiv \left.\frac{1}{T}\int_{t_0}^{t_0+T}dt~\mathcal{M}(t)\right|_{T\rightarrow \infty}$. For non-interacting spins, standard Rabi oscillations yield $\overline{\mathcal{M}}=-\frac{\Delta^2}{\Delta^2+\Omega_0^2}$, which decreases monotonically for $\Delta>0$. This behavior arises because large detuning suppresses the transitions between the $\ket{\uparrow}$ and $\ket{\downarrow}$ state. When interactions between the impurity and bosons are introduced, the $\ket{\downarrow}$ state can couple to the $\ket{\uparrow}$ state by exciting an additional phonon. This interaction mechanism leads to a qualitative difference in the steady-state magnetization for large $\Delta$, as shown in FIG. \ref{fig4}. The decrease for $k_n a=\pm 1$ at large $\Delta$ arises from the insufficient evolution time to fully reach the steady state. For $k_n a=-1$, the steady-state magnetization is well approximated by the zero-temperature ensemble of non-interacting spins, resulting in
\begin{equation}
\overline{\mathcal{M}}=\frac{\Delta}{\sqrt{\Delta^2+\Omega_0^2}}\ \ \ \ \ \ \text{for}\ a<0.
\end{equation}
In contrast, for $k_n a=1$, we observe the emergence of an additional peak near $E=E_a$. This feature is attributed to the contribution from attractive polarons, which significantly alters the behavior of the system. We propose a simple formula to capture both contributions, which reads 
\begin{equation}
\overline{\mathcal{M}}=\frac{\Delta}{\sqrt{\Delta^2+\Omega_0^2}}+ \frac{Z^{2}\Omega_0^2}{(\Delta-E_a)^2+Z\Omega_0^2}\ \ \ \text{for}\ a>0.
\end{equation}
Z is the quasi-particle residue of the second solution of \eqref{eqn:threelevel} at $\Delta=E_{a}$. These predictions are validated in FIG. \ref{fig4} using black dotted lines.
     \begin{figure}[t]
    \centering
    \includegraphics[width=0.99\linewidth]{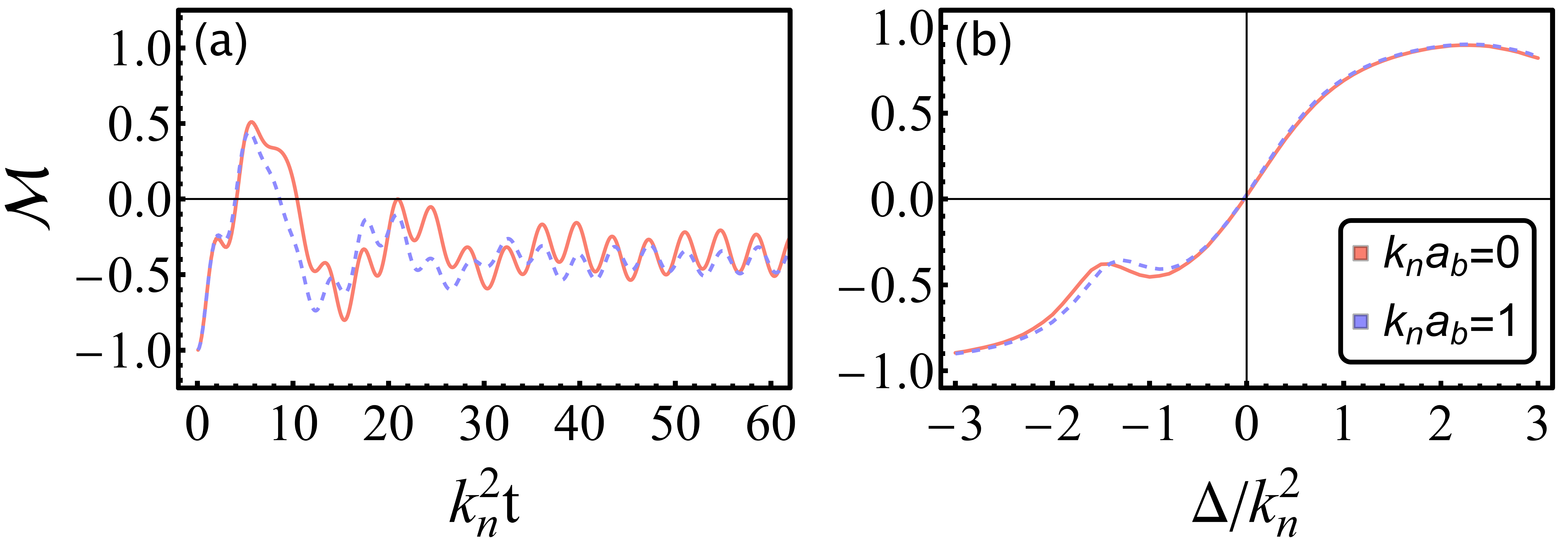}
    \caption{We present numerical results for the magnetization dynamics and the steady-state magnetization during Rabi oscillations for $k_na_b=0,1$, with fixed parameters $k_na=1$ and $\Omega_{0}/k_{n}^{2}=1$. In panel (a), we choose $\Delta=E_a\approx -1.34 k_n^2$. Both show a weak dependence on $k_n a_b$. }
    \label{fig5}
  \end{figure}

\textbf{ Interacting BEC.--}
Finally, we show that the above signatures of Rabi oscillations in driven Bose polarons remain valid even after introducing weak repulsive interactions, $k_na_b >0$. In this case, the solution of \eqref{eqn:evolution_eqn} becomes more involved and is deferred to the supplementary material \cite{SM}. The numerical results for the magnetization evolution $\mathcal{M}(t)$ are presented in FIG. \ref{fig5}, exhibiting the expected anomalous Rabi oscillation behavior for $k_na >0$ and $\Delta <0$. The magnetization of the steady-state also shows a weak dependence on $k_n a_b$. This demonstrates the stability of these intriguing dynamical behaviors.

  \emph{ \color{blue}Summary.--} 
In this work, we have explored the quantum dynamics of strongly driven Bose polarons in ultracold atomic gases, revealing rich non-equilibrium behavior through our trial wavefunction approach. Our analysis has uncovered anomalous Rabi oscillations and steady-state magnetization patterns that emerge due to the interplay between attractive and repulsive polaron branches, particularly in the BEC regime. These findings demonstrate how system-environment coupling fundamentally alters the dynamics of driven quasiparticles, bridging the gap between equilibrium polaron physics and non-equilibrium quantum dynamics and providing new insights for understanding and controlling quasiparticle behavior in strongly correlated systems. 

\vspace{5pt}
\textit{Acknowledgement.} 
would like to thank Xingyu Li and Hui Zhai for fruitful collaborations on related topics. This project is supported by the Innovation Program for Quantum Science and Technology 2024ZD0300101, the Shanghai Rising-Star Program under grant number 24QA2700300, and the NSFC under grant 12374477.

\bibliography{Polaron.bbl}

\ifarXiv


\section*{Supplementary material}

# Supplementary Material for Rabi Oscillations of Strongly Driven Bose Polaron

Zeyu Liu[1,2] and Pengfei Zhang[1,2,3,4,][*]

[1]*Department of Physics, Fudan University, Shanghai, 200438, China*
[2]*State Key Laboratory of Surface Physics, Fudan University, Shanghai, 200438, China*
[3]*Shanghai Qi Zhi Institute, AI Tower, Xuhui District, Shanghai 200232, China*
[4]*Hefei National Laboratory, Hefei 230088, China*

(Dated: April 18, 2025)

In this supplementary material, we present (1) Derivation of Eq. (7) in the main text; (2) Analytical results of Eq. (7) in the main text when $k_n a_b = 0$; (3) Solution of Eq. (7) in the main text with $0 < k_n a_b \ll 1$; (4) Quasiparticle properties without Rabi coupling; (5) Quasiparticle properties with Rabi coupling.

## I. DERIVATION OF EQ. (7) OF THE MAIN TEXT

We consider a single spin 1/2 impurity embedded in a cloud of Bose gas with weak repulsive interaction. Two internal states of the impurity are coupled by a radio-frequency field with Rabi coupling $\Omega_0$ and detuning $\Delta$ for $t > 0$. The total Hamiltonian of the system is given by

$$H = \sum_{k \neq 0} \gamma_k \beta_k^\dagger \beta_k + \sum_k \left( \epsilon_{i,k} f_{k,\uparrow}^\dagger f_{k,\uparrow} + (\epsilon_{i,k} + \Theta(t)\Delta) f_{k,\downarrow}^\dagger f_{k,\downarrow} + \frac{\Omega_0}{2} \Theta(t) \left( f_{k,\downarrow}^\dagger f_{k,\uparrow} + f_{k,\uparrow}^\dagger f_{k,\downarrow} \right) \right) + \frac{g}{V} \sum_{k,k',q} b_{k'}^\dagger f_{q-k',\uparrow}^\dagger f_{q-k,\uparrow} b_k. \quad (1)$$

Truncating the Hilbert space to at most one excitation, we write out the trial wavefunction [1].

$$|\psi(t)\rangle = \sum_\sigma \left( \psi_\sigma^{(0)}(t) f_{0,\sigma}^\dagger + \frac{1}{\sqrt{V}} \sum_{k \neq 0} \psi_{k,\sigma}^{(1)}(t) f_{-k,\sigma}^\dagger \beta_k^\dagger \right) |G\rangle. \quad (2)$$

$|G\rangle$ is the ground state without impurity and Bogoliubov quasiparticles. Then we can study the dynamics of the system using Schrödinger equation

$$i\partial_t |\psi(t)\rangle = H|\psi(t)\rangle. \quad (3)$$

Since we focus on the weakly interacting limit for the bosons, within the ground-state wavefunction $|G\rangle$, the annihilation and creation operators $b_0$ and $b_0^\dagger$ can be effectively replaced by $\sqrt{N_0}$. To proceed, we use the fact that our trial wavefunction includes at most one excitation. Consequently, terms involving $\beta_{k_1}\beta_{k_2}$ or $\beta_{k_1}^\dagger \beta_{k_2}^\dagger$ are negligible. This simplification leads to the following effective Hamiltonian [2].

$$\begin{aligned}H_{\text{eff}} = &\sum_{k \neq 0} \gamma_k \beta_k^\dagger \beta_k + \sum_k \left( (\epsilon_{i,k} + gn) f_{k,\uparrow}^\dagger f_{k,\uparrow} + (\epsilon_{i,k} + \Theta(t)\Delta) f_{k,\downarrow}^\dagger f_{k,\downarrow} + \frac{\Omega_0}{2} \Theta(t)(f_{k,\downarrow}^\dagger f_{k,\uparrow} + f_{k,\uparrow}^\dagger f_{k,\downarrow}) \right) \\ &+ \frac{g}{V} \left( \sqrt{N_0} \sum_{k \neq 0, q} \mathcal{R}_k f_{q+k,\uparrow}^\dagger f_{q,\uparrow} \left( \beta_k + \beta_{-k}^\dagger \right) + \sum_{k \neq 0, k' \neq 0, q} \mathcal{D}_{k,k'} f_{q+k-k',\uparrow}^\dagger f_{q,\uparrow} \beta_{k'}^\dagger \beta_k \right).\end{aligned} \quad (4)$$

Here, we have introduced

$$n = n_0 + n_1 = n_0 + \frac{1}{V} \sum_{k \neq 0} \langle G|\beta_k^\dagger \beta_k|G\rangle = n_0 + \frac{1}{V} \sum_{k \neq 0} v_k^2, \quad (5)$$

---

[*] PengfeiZhang.physics@gmail.com

and $\epsilon_k = \epsilon_{i,k} + \gamma_k$, $\mathcal{R}_k = u_k - v_k$, and $\mathcal{D}_{k,k'} = u_k u_{k'} + v_k v_{k'}$ for conciseness. Then it is straightforward to show that

$$\begin{aligned}
i\partial_t \psi_\uparrow^{(0)}(t) &= gn\psi_\uparrow^{(0)}(t) + \frac{\Omega_0}{2}\psi_\downarrow^{(0)}(t) + \frac{g}{V}\sqrt{n_0}\sum_k \mathcal{R}_k \psi_{k,\uparrow}^{(1)}(t), \\
i\partial_t \psi_{k,\uparrow}^{(1)}(t) &= \left(\epsilon_k + gn\right)\psi_{k,\uparrow}^{(1)}(t) + \frac{\Omega_0}{2}\psi_{k,\downarrow}^{(1)}(t) + g\sqrt{n_0}\mathcal{R}_k \psi_\uparrow^{(0)}(t) + \frac{g}{V}\sum_{k'} \mathcal{D}_{k,k'} \psi_{k',\uparrow}^{(1)}(t), \\
i\partial_t \psi_\downarrow^{(0)}(t) &= \Delta\psi_\downarrow^{(0)}(t) + \frac{\Omega_0}{2}\psi_\uparrow^{(0)}(t), \\
i\partial_t \psi_{k,\downarrow}^{(1)}(t) &= \left(\epsilon_k + \Delta\right)\psi_{k,\downarrow}^{(1)}(t) + \frac{\Omega_0}{2}\psi_{k,\uparrow}^{(1)}(t).
\end{aligned} \qquad (6)$$

This reproduces Eq. (7) of the main text, provided that we do not distinguish between $n_0$ and $n$. The initial condition is $\psi_\downarrow^{(0)}(0) = 1$, while all other components of the wavefunction vanish. The dynamics of strongly driven Bose polarons is determined by this set of equations.

## II. ANALYTICAL RESULTS OF EQ. (7) OF THE MAIN TEXT WHEN $k_n a_b = 0$

Now we consider the simpler case where $k_n a_b = 0$. After performing the Laplace transform $F(s) \equiv \int_0^\infty e^{-st} f(t) dt$, Eq. (6) becomes

$$\begin{aligned}
is\psi_\uparrow^{(0)}(s) &= \frac{\Omega_0}{2}\psi_\downarrow^{(0)}(s) + \sqrt{n_0}\chi(s), \\
is\psi_{k,\uparrow}^{(1)}(s) &= \left(\epsilon_k + gn_0\right)\psi_{k,\uparrow}^{(1)}(s) + \frac{\Omega_0}{2}\psi_{k,\downarrow}^{(1)}(s) + \chi(s), \\
is\psi_\downarrow^{(0)}(s) &= \Delta\psi_\downarrow^{(0)}(s) + \frac{\Omega_0}{2}\psi_\uparrow^{(0)}(s) + i, \\
is\psi_{k,\downarrow}^{(1)}(s) &= \left(\epsilon_k + \Delta\right)\psi_{k,\downarrow}^{(1)}(s) + \frac{\Omega_0}{2}\psi_{k,\uparrow}^{(1)}(s), \\
\chi(s) &\equiv g\sqrt{n_0}\psi_\uparrow^{(0)}(s) + \frac{g}{V}\sum_{k\neq 0}\psi_{k,\uparrow}^{(1)}(s).
\end{aligned} \qquad (7)$$

We can express the components in the wavefunction via $\chi(s)$. Concretely speaking, we have

$$\begin{aligned}
\psi_\uparrow^{(0)}(s) &= \frac{\frac{i\Omega_0}{2} + (is - \Delta)\sqrt{n_0}\chi(s)}{D_0}, \\
\psi_{k,\uparrow}^{(1)}(s) &= \frac{(is - \epsilon_k - \Delta)\chi(s)}{D_{k,g}},
\end{aligned} \qquad (8)$$

where we have defined $D_0 = is(is - \Delta) - (\Omega_0/2)^2$ and $D_{k,g} = (is - \epsilon_k - gn_0)(is - \epsilon_k - \Delta) - (\Omega_0/2)^2$. Combining the definition of $\chi(s)$ and Eq. (8), we have

$$\left(\frac{1}{g} - \frac{(is - \Delta)n_0}{D_0} - \frac{1}{V}\sum_{k\neq 0}\frac{is - \epsilon_k - \Delta}{D_{k,g}}\right)\chi(s) = \frac{i\Omega_0\sqrt{n_0}}{2D_0}. \qquad (9)$$

Define

$$\begin{aligned}
\Pi(is) &\equiv \frac{1}{g} - \frac{1}{V}\sum_{k\neq 0}\frac{is - \epsilon_k - \Delta}{D_{k,g}} \\
&= \frac{m}{2\pi a} - \frac{1}{V}\sum_k \frac{2m}{k^2} - \frac{1}{V}\sum_{k\neq 0}\frac{is - \epsilon_k - \Delta}{D_k} \\
&= \frac{m}{2\pi a} - \frac{m^{3/2}}{2\pi\sqrt{2}}\left(\left(1 + \frac{\Delta}{\sqrt{\Omega_0^2 + \Delta^2}}\right)\sqrt{-is + \frac{\Delta}{2} - \frac{\sqrt{\Omega_0^2 + \Delta^2}}{2}} + \left(1 - \frac{\Delta}{\sqrt{\Omega_0^2 + \Delta^2}}\right)\sqrt{-is + \frac{\Delta}{2} + \frac{\sqrt{\Omega_0^2 + \Delta^2}}{2}}\right),
\end{aligned} \qquad (10)$$

where $D_k = (is - \epsilon_k)(is - \epsilon_k - \Delta) - (\Omega_0/2)^2$, and we have used the renormalization relation $\frac{1}{g} = \frac{m}{2\pi a} - \frac{1}{V}\sum_k \frac{2m}{k^2}$. Then it is straightforward to show that

$$\begin{aligned}
\psi_\uparrow^{(0)}(s) &= \frac{i(\Omega_0/2)}{(is - \Delta)(is - n_0\Pi^{-1}(is)) - (\Omega_0/2)^2}, \\
\psi_\downarrow^{(0)}(s) &= \frac{i}{is - \Delta - (\Omega_0/2)^2 \left(is - n_0\Pi^{-1}(is)\right)^{-1}}, \\
\psi_{k,\uparrow}^{(1)}(s) &= \frac{i\sqrt{n_0}\,(\Omega_0/2)\,(is - \epsilon_k - \Delta)}{D_k\bigl(D_0\Pi(is) - (is - \Delta)n_0\bigr)}, \\
\psi_{k,\downarrow}^{(1)}(s) &= \frac{i\sqrt{n_0}\,(\Omega_0/2)^2}{D_k\bigl(D_0\Pi(is) - (is - \Delta)n_0\bigr)}.
\end{aligned} \tag{11}$$

These results provide the exact expressions for each element of the wavefunction. However, we cannot perform the inverse Laplace transform analytically.

### III.  SOLUTION OF EQ. (7) OF THE MAIN TEXT WHEN $0 < k_n a_b \ll 1$

Now we consider the case where the bosons have weak repulsive interactions. We start with the definitions

$$\begin{aligned}
\mathcal{U}(s) &\equiv \frac{g}{V}\sum_{k\neq 0} u_k \psi_{k,\uparrow}^{(1)}(s), \\
\mathcal{V}(s) &\equiv \frac{g}{V}\sum_{k\neq 0} v_k \psi_{k,\uparrow}^{(1)}(s).
\end{aligned} \tag{12}$$

Then, we can rewrite Eq. (6) as

$$\begin{aligned}
is\psi_\uparrow^{(0)}(s) &= gn\psi_\uparrow^{(0)}(s) + \frac{\Omega_0}{2}\psi_\downarrow^{(0)}(s) + \sqrt{n_0}\,\mathcal{U}(s) - \sqrt{n_0}\,\mathcal{V}(s), \\
is\psi_{k,\uparrow}^{(1)}(s) &= (\epsilon_k + gn)\psi_{k,\uparrow}^{(1)}(s) + \frac{\Omega_0}{2}\psi_{k,\downarrow}^{(1)}(s) + g\sqrt{n_0}\,\mathcal{R}_k\psi_\uparrow^{(0)}(s) + u_k\mathcal{U}(s) + v_k\mathcal{V}(s), \\
is\psi_\downarrow^{(0)}(s) &= \Delta\psi_\downarrow^{(0)}(s) + \frac{\Omega_0}{2}\psi_\uparrow^{(0)}(s) + i, \\
is\psi_{k,\downarrow}^{(1)}(s) &= (\epsilon_k + \Delta)\psi_{k,\downarrow}^{(1)}(s) + \frac{\Omega_0}{2}\psi_{k,\uparrow}^{(1)}(s).
\end{aligned} \tag{13}$$

Now we can express the components of the wavefunction in terms of $\mathcal{U}(s)$ and $\mathcal{V}(s)$. Explicitly, we obtain

$$\begin{aligned}
\psi_\uparrow^{(0)}(s) &= \frac{1}{D_0}\left((is - \Delta)\,gn\psi_\uparrow^{(0)}(s) + \frac{i\Omega_0}{2} + \sqrt{n_0}\,(is - \Delta)(\mathcal{U}(s) - \mathcal{V}(s))\right), \\
\psi_{k,\uparrow}^{(1)}(s) &= \frac{is - \epsilon_k - \Delta}{D_{k,g}}\left(g\sqrt{n_0}\,\mathcal{R}_k\psi_\uparrow^{(0)}(s) + u_k\mathcal{U}(s) + v_k\mathcal{V}(s)\right),
\end{aligned} \tag{14}$$

Here we have defined $D_0 = is(is - \Delta) - (\Omega_0/2)^2$ and $D_{k,g} = (is - \epsilon_k - gn)(is - \epsilon_k - \Delta) - (\Omega_0/2)^2$. By combining the definitions of $\mathcal{U}(s)$ and $\mathcal{V}(s)$ and Eq. (14), we have

$$\begin{aligned}
\mathcal{U}(s) &= g^2\sqrt{n_0}\,\psi_\uparrow^{(0)}(s)\bigl(\Pi_{uu}(s) - \Pi_{uv}(s)\bigr) + g\Pi_{uu}(s)\mathcal{U}(s) + g\Pi_{uv}(s)\mathcal{V}(s), \\
\mathcal{V}(s) &= g^2\sqrt{n_0}\,\psi_\uparrow^{(0)}(s)\bigl(\Pi_{uv}(s) - \Pi_{vv}(s)\bigr) + g\Pi_{uv}(s)\mathcal{U}(s) + g\Pi_{vv}(s)\mathcal{V}(s),
\end{aligned} \tag{15}$$

where we have introduced

$$\begin{aligned}
\Pi_{uu}(s) &= \frac{1}{V}\sum_{k\neq 0}\frac{(is - \epsilon_k - \Delta)u_k^2}{D_{k,g}}, \\
\Pi_{uv}(s) &= \frac{1}{V}\sum_{k\neq 0}\frac{(is - \epsilon_k - \Delta)u_k v_k}{D_{k,g}}, \\
\Pi_{vv}(s) &= \frac{1}{V}\sum_{k\neq 0}\frac{(is - \epsilon_k - \Delta)v_k^2}{D_{k,g}}.
\end{aligned} \tag{16}$$

To proceed, we define

$$\hat{\Pi}(s) = \begin{pmatrix} \Pi_{uu}(s) & \Pi_{uv}(s) \\ \Pi_{uv}(s) & \Pi_{vv}(s) \end{pmatrix}. \tag{17}$$

Now we can rewrite Eq. (15) as

$$\begin{pmatrix} \mathcal{U}(s) \\ \mathcal{V}(s) \end{pmatrix} = g\sqrt{n_0}\psi_\uparrow^{(0)}(s) \frac{\hat{\Pi}(s)}{\frac{1}{g}\hat{I} - \hat{\Pi}(s)} \begin{pmatrix} 1 \\ -1 \end{pmatrix}. \tag{18}$$

In Eq. (17), only $\Pi_{uu}(s)$ diverges, so we regularize it using

$$\frac{1}{g} = \frac{m}{2\pi a} - \frac{m\Lambda}{\pi^2}. \tag{19}$$

$\Lambda$ is the ultraviolet cutoff. Similarly, we redefine $\Pi_{uu}(s)$ as

$$\Pi_{uu}(s) = \Pi_{uu}^{\text{reg}}(s) - \frac{m\Lambda}{\pi^2}, \tag{20}$$

where $\Pi_{uu}^{\text{reg}}(s)$ converges. Using Eq. (19) and Eq. (20) and performing Taylor expansion with repsect to $\Lambda^{-1}$, we have

$$\begin{aligned} g &= -\frac{\pi^2}{m}\frac{1}{\Lambda} + O(\frac{1}{\Lambda^2}), \\ \frac{g\hat{\Pi}(s)}{\frac{1}{g}\hat{I} - \hat{\Pi}(s)} &= \begin{pmatrix} \left(\frac{1}{g} - \Pi_{uu}(s)\right)^{-1} & 0 \\ 0 & 0 \end{pmatrix} + O(\frac{1}{\Lambda}). \end{aligned} \tag{21}$$

Then it is straightforward to show that

$$\begin{aligned}
\mathcal{A}(is) &\equiv \frac{1}{g} - \Pi_{uu}(s) = \frac{m}{2\pi a} - \frac{1}{V}\sum_k \left(\frac{2m}{k^2} + \frac{(is - \epsilon_k - \Delta)u_k^2}{D_k}\right), \\
\psi_\uparrow^{(0)}(s) &= \frac{i(\Omega_0/2)}{(is-\Delta)(is - n_0\mathcal{A}^{-1}(is)) - (\Omega_0/2)^2}, \\
\psi_\downarrow^{(0)}(s) &= \frac{i}{is - \Delta - (\Omega_0/2)^2\left(is - n_0\mathcal{A}^{-1}(is)\right)^{-1}}, \\
\psi_{k,\uparrow}^{(1)}(s) &= \frac{i\sqrt{n_0}u_k(\Omega_0/2)(is - \epsilon_k - \Delta)}{D_k\bigl(D_0\mathcal{A}(is) - (is-\Delta)n_0\bigr)}, \\
\psi_{k,\downarrow}^{(1)}(s) &= \frac{i\sqrt{n_0}u_k(\Omega_0/2)^2}{D_k\bigl(D_0\mathcal{A}(is) - (is-\Delta)n_0\bigr)}.
\end{aligned} \tag{22}$$

This is the generalization of Eq. (11) in the weakly interacting limit. We cannot solve it analytically.

## IV. QUASIPARTICLE PROPERTIES WITHOUT THE RABI COUPLING

In this section, we study the energy of attractive and repulsive polarons without Rabi coupling in the non-interacting regime. We start with the Hamiltonian

$$H = \sum_k \epsilon_{b,k} b_k^\dagger b_k + \sum_k \epsilon_{i,k} f_k^\dagger f_k + \frac{g}{V}\sum_{k,k',q} b_{k'}^\dagger f_{q-k'}^\dagger f_{q-k} b_k. \tag{23}$$

Now we write down the trial wavefunction

$$|\psi\rangle = \left(\psi_0 f_0^\dagger + \frac{1}{\sqrt{V}}\sum_{k\neq 0} \psi_k f_{-k}^\dagger b_k^\dagger\right)|G\rangle. \tag{24}$$

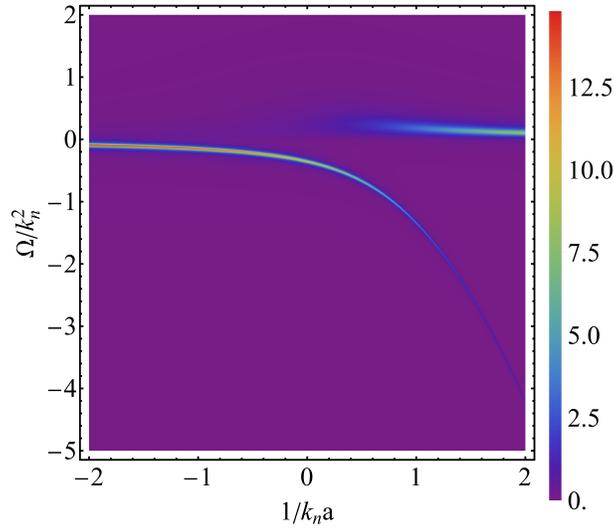

FIG. 1. We present the density plot for the spectral function (without rabi coupling) with $k_n a_b = 0$.

Under this wavefunction, we have

$$\rho_k \equiv \psi_k/\psi_0$$
$$E = gn_0 + \frac{g}{V}\sqrt{n_0}\sum_{k\neq 0}\rho_k \quad (25)$$
$$(E - \epsilon_k - gn_0)\rho_k = g\sqrt{n_0} + \frac{g}{V}\sum_{k\neq 0}\rho_k.$$

Here $\epsilon_k = \epsilon_{b,k} + \epsilon_{i,k}$. Then we have

$$E = n_0\left(\frac{1}{g} - \frac{1}{V}\sum_{k\neq 0}\frac{1}{E-\epsilon_k-gn_0}\right)^{-1} = n_0\left(\frac{m}{2\pi a} - \frac{1}{V}\sum_{k}\frac{1}{\epsilon_k} - \frac{1}{V}\sum_{k\neq 0}\frac{1}{E-\epsilon_k}\right)^{-1}. \quad (26)$$

The self-energy then reads

$$\Sigma(\Omega + i0^+) = n_0\left(\frac{m}{2\pi a} - \frac{1}{V}\sum_{k}\frac{1}{\epsilon_k} - \frac{1}{V}\sum_{k\neq 0}\frac{1}{\Omega + i0^+ - \epsilon_k}\right)^{-1}$$
$$= n_0\left(\frac{m}{2\pi a} - \frac{m^{3/2}\sqrt{-\Omega - i0^+}}{\pi\sqrt{2}}\right)^{-1}. \quad (27)$$

The quasiparticle energy is defined as the real solution of

$$\Omega = \text{Re}(\Sigma(\Omega)). \quad (28)$$

If $\Omega < 0$, we have

$$\Sigma(\Omega) = n_0\left(\frac{m}{2\pi a} - \frac{m^{3/2}}{\pi\sqrt{2}}\sqrt{-\Omega}\right)^{-1}. \quad (29)$$

There is only one solution of Eq. (28) which reads

$$E_a = \frac{-1}{18m}\left(a^{-1} + \sqrt[3]{a^{-3} + 54\pi n_0 + 6\sqrt{3\pi n_0(a^{-3} + 27\pi n_0)}} + \frac{a^{-2}}{\sqrt[3]{a^{-3} + 54\pi n_0 + 6\sqrt{3\pi n_0(a^{-3} + 27\pi n_0)}}}\right)^2. \quad (30)$$

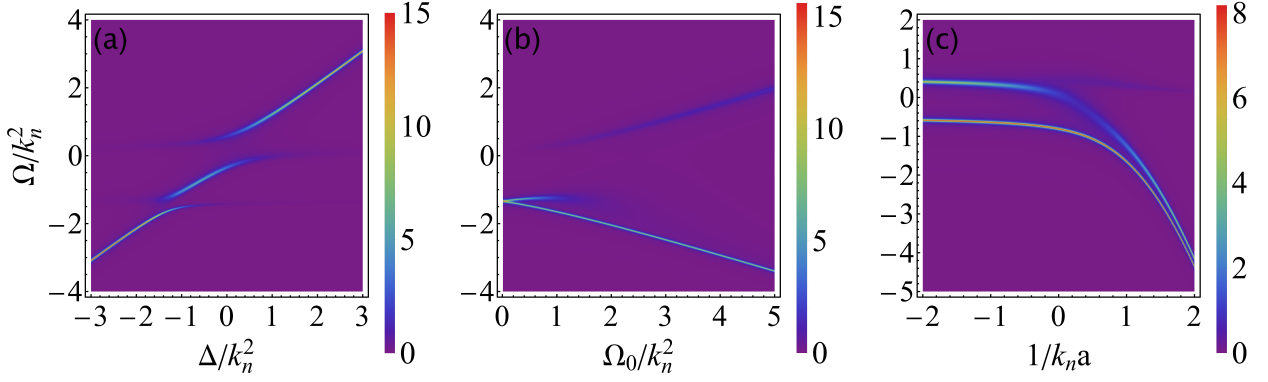

FIG. 2. We present the density plot for the impurity spectral function. In all plots, we set $k_n a_b = 0$. In panel (a), we set $\Omega_0/k_n^2 = 1$ and $k_n a = 1$. In panel (b), we set $k_n a = 1$ and $\Delta = E_a$. In panel (c), we set $\Omega_0/k_n^2 = 1$, and $\Delta = E_a$.

where $Z_a$ is the quasiparticle residue of attractive polaron defined as

$$Z_a = \left(\frac{1}{1 - \partial_\Omega \Sigma(\Omega)}\right)_{\Omega = E_a}. \tag{31}$$

For $\Omega > 0$, we instead find

$$\Sigma(\Omega + i0^+) = n_0 \left(\frac{m}{2\pi a} + i\frac{m^{3/2}}{\pi \sqrt{2}} \sqrt{\Omega}\right)^{-1}. \tag{32}$$

There is no positive solution of Eq. (28) when $a < 0$ and only one positive solution when $a > 0$ which is given by

$$E_r = \frac{-1 + \sqrt{1 + 16\pi n_0 a^3}}{4 m a^2}. \tag{33}$$

Finally, the spectral function is defined as

$$A(\Omega) \equiv -\frac{1}{\pi} \text{Im}\left(\frac{1}{\Omega + i0^+ - \Sigma(\Omega + i0^+)}\right) = Z_a \delta(\Omega - E_a) - \frac{\Theta(\Omega)}{\pi} \text{Im}\left(\frac{1}{\Omega - n\left(\frac{m}{2\pi a} + i\frac{m^{3/2}}{\sqrt{2\pi}} \sqrt{\Omega}\right)^{-1}}\right). \tag{34}$$

This completes our discussion of quasiparticle properties without Rabi drive. The numerical result for the spectral function is shown in Fig. 1.

## V. QUASIPARTICLE PROPERTIES WITH THE RABI COUPLING

Next, similar to Eq. (25), for Bose polaron with Rabi drive, we have

$$\begin{aligned}
E \rho_0 &= \frac{\Omega_0}{2} + \sqrt{n_0} \chi, \\
E &= \Delta + \frac{\Omega_0}{2} \rho_0, \\
(E - \epsilon_k - \Delta) \rho_{k,\downarrow} &= \frac{\Omega_0}{2} \rho_{k,\uparrow}, \\
(E - \epsilon_k - gn) \rho_{k,\uparrow} &= \frac{\Omega_0}{2} \rho_{k,\downarrow} + \chi, \\
\chi &\equiv g \sqrt{n_0} \rho_0 + \frac{g}{V} \sum_{k \neq 0} \rho_{k,\uparrow},
\end{aligned} \tag{35}$$

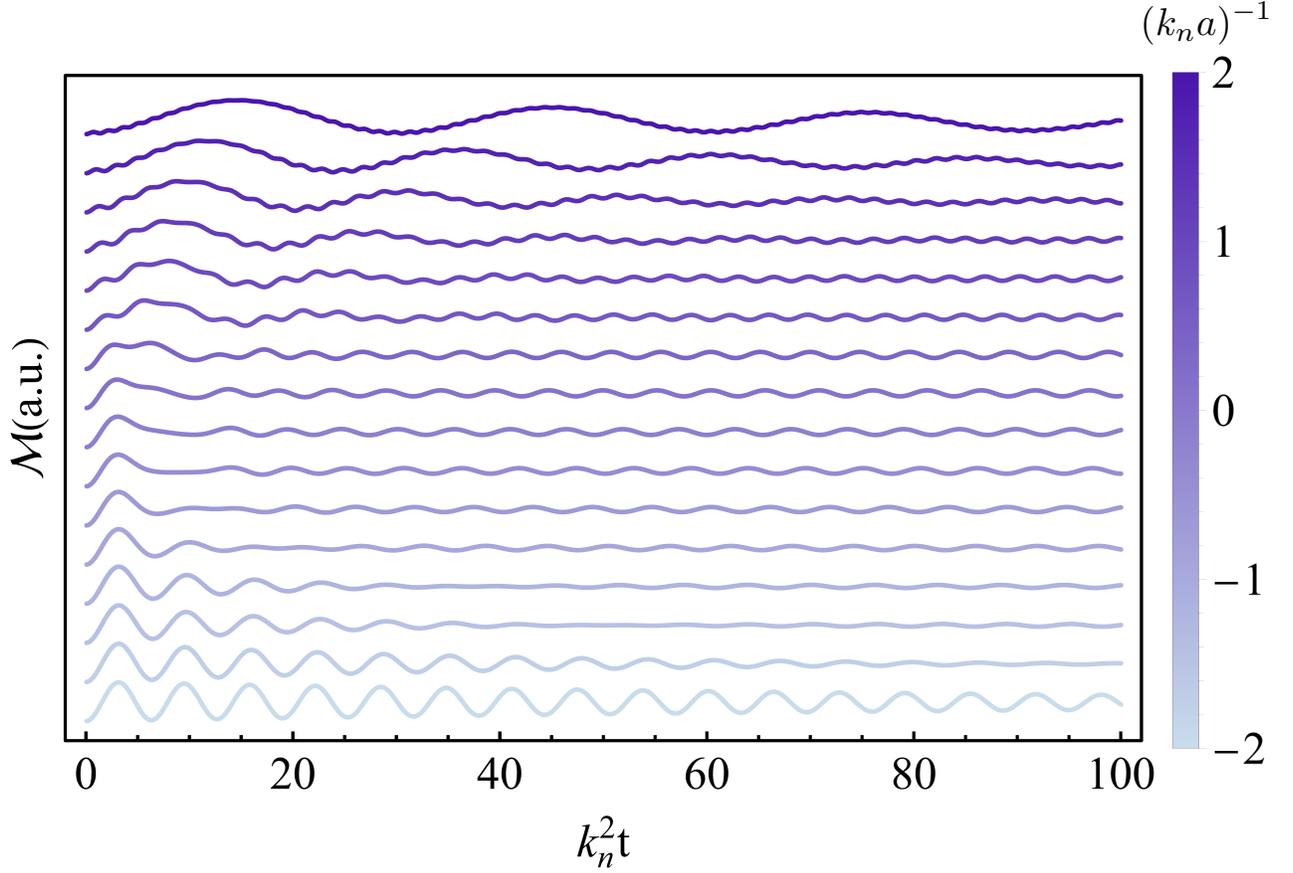

FIG. 3. We present numerical results for the magnetization dynamics during Rabi oscillations with $k_n a_b = 0$, $\Omega_0/k_n^2 = 1$, $\Delta = E_a$, and $(k_n a)^{-1} \in \{-2, 2\}$. The curves are artificially shifted vertically for clarity.

where $\rho_0 = \psi_{0,\uparrow}/\psi_{0,\downarrow}$, $\rho_{k,\uparrow} = \psi_{k,\uparrow}/\psi_{0,\downarrow}$, and $\rho_{k,\downarrow} = \psi_{k,\downarrow}/\psi_{0,\downarrow}$. Introduce $\Pi(E)$ as

$$\Pi(E) = \frac{1}{g} - \frac{1}{V} \sum_{k \neq 0} \frac{E - \epsilon_k - \Delta}{D_k}, \tag{36}$$

where $D_k = (E - \epsilon_k)(E - \epsilon_k - \Delta) - \left(\frac{\Omega_0}{2}\right)^2$. Then it is straightforward to show that

$$E = \Delta + \frac{(\Omega_0/2)^2}{E - n_0 \Pi^{-1}(E)}. \tag{37}$$

This equation should be interpreted as looking for the pole of the impurity green's function

$$G_\downarrow^R(\Omega) = \frac{1}{\Omega - \Delta - \frac{(\Omega_0/2)^2}{\Omega - n_0 \Pi^{-1}(\Omega)}}. \tag{38}$$

Finally, we can compute the spectral function using

$$A_\downarrow(\Omega) = -\frac{1}{\pi} \mathrm{Im}\left(G_\downarrow^R(\Omega + i0^+)\right). \tag{39}$$

Numerical results for the spectral function are shown in Fig. 2. Next, we present the perturbative analysis of $\Omega_0$. In the limit of $\Omega_0 \to 0$, we have $n_0 \Pi^{-1}(E) = \Sigma(E)$.

For $k_n a < 0$, there are two branches of excitations. Recall that the polaron energy $E_a$ and the quasiparticle residue $Z_a$ are determined by solving $E_a = \Sigma(E_a)$ and $Z_a = (1 - \partial_\Omega \Sigma(E_a))^{-1}$. So we have

$$\frac{1}{\Omega - n_0 \Pi^{-1}(\Omega)} \approx \frac{Z_a}{\Omega - E_a}. \tag{40}$$

7

The energy of the quasiparticle with Rabi coupling is determined by the solution of $\text{Re}\left[G_\downarrow^{-1}(\Omega)\right] = 0$. Under these approximations, the excitation energies $\Omega^*$ are given by

$$\Omega^* = \frac{E_a + \Delta \pm \sqrt{(E_a - \Delta)^2 + Z_a \Omega_0^2}}{2} \quad \text{for } a < 0. \tag{41}$$

For comparison, there are three branches of excitations for $k_n a > 0$ due to the coexistence of attractive and repulsive polarons, which serves as the underlying physical origin of anomalous Rabi oscillation (Fig. 3). We assume that the spectral function without Rabi coupling can be expressed as

$$A(\Omega) = Z_a \delta(\Omega - E_a) + (1 - Z_a)\delta(\Omega - E_r), \tag{42}$$

which satisfies

$$\int_{-\infty}^{\infty} d\Omega\, A(\Omega) = 1. \tag{43}$$

This leads to

$$\frac{1}{\Omega - n_0 \Pi^{-1}(\Omega)} \approx \frac{Z_a}{\Omega - E_a} + \frac{1 - Z_a}{\Omega - E_r}. \tag{44}$$

Finally, we come to the equation for excitation energies

$$(\Omega^* - \Delta) = \frac{(\Omega_0/2)^2 Z_a}{\Omega^* - E_a} + \frac{(\Omega_0/2)^2(1 - Z_a)}{\Omega^* - E_r} \quad \text{for } a > 0. \tag{45}$$

These solutions match the numerical results with good accuracy.

---


\fi

\end{document}